%% file: paper.tex
\documentclass[sigconf, nonacm]{acmart}
\usepackage{xspace}
\usepackage{cleveref}
\usepackage{subcaption}
\usepackage{graphicx}
\usepackage{microtype}
\usepackage{hyperref}
\usepackage{mathtools}
\usepackage{amsthm}
\usepackage{listings}

\usepackage{multirow}
\usepackage{siunitx}
\usepackage{framed}
\usepackage{xcolor}
\usepackage{dcolumn}
\usepackage{colortbl}

\definecolor{mygreen}{HTML}{1bc304}

\usepackage{pifont}

\usepackage{ifthen}

\usepackage[ruled,vlined,linesnumbered,inoutnumbered,noend]{algorithm2e}

\makeatletter
\renewcommand\@makefnmark{\hbox{\@textsuperscript{\normalfont\color{blue}\@thefnmark}}}
\makeatother

\usepackage[breakable, skins,xparse]{tcolorbox}

\input{defs}

\newcommand{\appendixsite}{our Supplement~\cite{supplweb}}

\hypersetup{
    colorlinks = true,
    linkcolor = blue,
    filecolor = blue,      
    urlcolor = blue,
   citecolor = blue,
}

\newboolean{changed}
\setboolean{changed}{true}
\newcommand{\changed}[2]{\ifthenelse{\boolean{changed}}{\color{red}#1\color{black}}{\color{black}#2\color{black}}}

\newboolean{showBothMode}
\setboolean{showBothMode}{false} 

\newcommand{\changedFinalRevision}[2]{%
  \ifthenelse{\boolean{showBothMode}}%
    {{\color{red}\sout{#2}}{\color{cyan}#1}}
    {{\color{black}#1}}
}
\newcommand{\changedFinalRevisiontwo}[2]{%
  \ifthenelse{\boolean{showBothMode}}%
    {{\color{red}\sout{#2}}{\color{blue}#1}}
    {{\color{black}#1}}
}

\newboolean{cut}
\setboolean{cut}{true}
\newcommand{\cut}[2]{\ifthenelse{\boolean{cut}}{\color{yellow}#1\color{black}}{\color{black}#2\color{black}}}

\definecolor{darkblue}{rgb}{0,0,.75}

\input{listing_defs}

\theoremstyle{definition}
\newtheorem{definition}{Definition}[section]

\begin{document}

\title{Multi-Location Software Model Completion }%

\author{Alisa Welter}
\email{welter@uni-saarland.de}
\affiliation{%
  \institution{Saarland University}
  \city{Saarbr\"ucken}
  \country{Germany}
}

\author{Christof Tinnes}
\email{christof.tinnes@siemens.com}
\affiliation{%
  \institution{Saarland University}
  \city{Saarbr\"ucken}
  \country{Germany}
}

\author{Sven Apel}
\email{apel@uni-saarland.de}
\affiliation{%
  \institution{Saarland University}
  \city{Saarbr\"ucken}
  \country{Germany}
}

\renewcommand{\shortauthors}{Welter, Tinnes, and Apel}

\input{Content/0_abstract}

\maketitle              

\input{Content/1_introduction}

\input{Content/2_relatedwork}

\input{Content/3_background}

\input{Content/5_approach}

\input{Content/6_experiments}

\input{Content/8_Dicussion}

\input{Content/7_conclusion}

\def\bibfont{\footnotesize}
\bibliographystyle{ACM-Reference-Format}
\bibliography{bib}
\end{document}

%% file: defs.tex
\newcommand{\Random}{\textit{random selection}}
\newcommand{\Semantic}{\textit{semantic similarity}}
\newcommand{\Historical}{\textit{historical co-change frequency}}

\newcommand{\nameapproach}{\textsc{Next\-Focus}\xspace}

\newcommand{\nameapproachfull}{global embedding-based next focus predictor\xspace}

\newcommand{\nametinnes}{\textsc{RaMc}\xspace}
\newcommand{\nametinnesbetter}{\textsc{RaMc$^{\prime}$}\xspace}

\newcommand{\CorrectnessValueNextFocus}{63.94\%\xspace}
\newcommand{\CorrectnessValueNextFocusTinnes}{30.18\%\xspace}

\newcommand{\StructureImproved}{40.68\%\xspace}


%% file: listing_defs.tex
\definecolor{eminence}{RGB}{108,48,130}

\lstnewenvironment{PseudoCode}[1][]
{\lstset{language=Matlab,basicstyle=\scriptsize, keywordstyle=\color{darkblue},numbers=left,xleftmargin=.04\textwidth,#1}}
{}

%% file: Content/0_abstract.tex
\begin{abstract}

In model-driven engineering and beyond, software models are key development artifacts. In practice, they often grow to substantial size and complexity, undergoing thousands of modifications over time due to evolution, refactoring, and maintenance. The rise of AI has sparked interest in how software modeling activities can be automated. Recently, LLM-based approaches for software model completion have been proposed, however, the state of the art supports only single-location model completion by predicting changes at a specific location. Going beyond, we aim to bridge the gap toward handling coordinated changes that span multiple locations across large, complex models.
Specifically, we propose a novel \nameapproachfull, \nameapproach, which is capable of multi-location model completion for the first time. The predictor consists of a neural network with an attention mechanism that is trained on historical software model evolution data. Starting from an existing change, it predicts further model elements to change, potentially spanning multiple parts of the model.
 We evaluate our approach on multi-location model changes that have actually been performed by developers in real-world projects.
\nameapproach{} achieves promising results for multi-location model completion, even when changes are heavily spread across the model. It \changedFinalRevision{achieves}{archives }an average Precision@\(k\) score of 0.98 for \(k \leq 10\), significantly outperforming the three baseline approaches.

\end{abstract}


%% file: Content/1_introduction.tex
\section{Introduction}\label{sec:introduction}

In model-driven engineering and beyond, software models help bridge the gap between the problem domain and the implementation domain by offering multiple levels and types of abstraction, thereby reducing overall system complexity~\cite{france2007model}.
In practice, for example, in industrial automation and automotive engineering, where a substantial fraction of code is generated from models, these software models can become very large and complex~\cite{tinnes2022sometimes}. 
For example, a single subsystem may undergo thousands of individual modifications when transitioning from the main development branch to customized versions~\cite{tinnes2022sometimes}.
In general, changes tend to quickly grow and cut across the model~\cite{kretschmer2021consistent,
briand2003impact,
ohrndorf2021b,tinnes2023mining}. Even a \textit{single, local change} may require complex adjustments in other parts to preserve or correctly extend the system’s functionality and semantics. 
This makes maintaining and evolving software models tedious, time-consuming, and error-prone~\cite{tinnes2023mining, tinnes2024leveraging}.

To address these challenges, initial steps have been taken to automate software model evolution, powered by the rise of AI.
One area of focus is \emph{software model completion}, where a (partial) software model is provided and a tool suggests changes to the software model. Before the advent of large language models (LLMs), previous work often relied on predefined sets of model completion patterns and used (semi-) automated, rule-based techniques to recommend completions for software models~\cite{kuschke2017rapmod, kuschke2013recommending, maeder2021, li2013efficient, deng2016recommendation, kogel2016automatic,kogel2017recommender}. However, this approach is limited, as each new project-specific pattern requires defining additional edit rules. Specifying edit rules typically demands expertise in both the specification and domain-specific languages, and their evolution over time -- such as through metamodel changes -- adds further complexity.

Advancements in AI have 
opened up new possibilities for software modeling~\cite{camara2023assessment, eisenreich2024requirements,chaaben2023towards, tinnes2024leveraging, chaaben2024utility}. 
Recently, LLMs from the GPT family have been used successfully for model completion~\cite{chaaben2023towards, tinnes2024leveraging, chaaben2024utility}.
In particular, the general inference capabilities of LLMs are useful for handling domain concepts with few or no similar examples, which is common in the modeling domain. They have been shown to be effective at dealing with verbose and noisy textual components found in domain-specific modeling data in industry, making them valuable in scenarios where other approaches fall short~\cite{tinnes2024leveraging}.


Despite considerable progress, existing LLM-based approaches are typically limited to \textit{single-location changes}. That is, they modify, extend, or add one or more elements that are directly connected to each other at a single location in the software model~\cite{tinnes2024leveraging}. In practice, however, a single local change may require adjustments in other parts of the model. In general, bug fixes and feature additions may affect many different locations~\cite{rashid2011aspect,apel2011language}. We call these changes \textit{multi-location software model changes}. 
Multi-location changes are particularly challenging to manage, as dependencies across the model can be easily overlooked, a problem that is well understood in the realm of code~\cite{kiczales1997aspect, eaddy2008crosscutting, apel2013feature}. Applying them correctly is often error-prone, time-consuming and requires substantial domain knowledge to understand what needs to be changed and where -- especially given the sheer size of real-world software models.

Addressing this problem, we propose an approach for \textit{multi-location model completion}, that implicitly learns multi-location co-change patterns from data. Given a single-location model edit (by the user), a \nameapproachfull, \nameapproach, suggests further locations anywhere in the model to be edited as well, based on similar patterns observed in the data. Technically, \nameapproach rests on a neural network with an attention layer that, given historical pairs of co-changed nodes as training data, ranks them and suggests them to the user. 

For evaluation, we investigate the performance of \nameapproach for multi-location model completion on a real-world dataset containing \changedFinalRevision{32}{41} projects with multi-location changes that were actually performed by modelers in a real-world scenario. For this purpose, we rely on standard recommendation metrics, in particular, Precision@\(k\). We found that \nameapproach{} achieves an average score of 0.98 over all \(k \in \{1, \ldots, 10\}\), significantly outperforming three baselines. Notably, \nameapproach performs well even when changes are spread across a software model. A manual investigation revealed patterns that worked well and those that did not: we observed high predictive performance, especially in structured, frequently recurring patterns -- such as changes involving the renaming or replacement of existing types, but also the introduction of entirely new domain concepts. On the other hand, \nameapproach{} (and baselines) struggle with some cases, e.g., when the hierarchy of modeling elements was changed. \changedFinalRevision{Finally, we evaluate \nameapproach in an iterative, multi-location completion setting by combining it with state of the art, LLM-based, single-location model completion~\cite{tinnes2024leveraging}.}{}


In summary, we make the following contributions: 
\vskip -2ex
\begin{itemize}
    \item We define the notion of multi-location model completion based on single-location model completion~\cite{tinnes2024leveraging}.
    \item We propose a \nameapproachfull{} for \textit{multi-location model completion}, \nameapproach, that predicts new change locations based on historical data.
   \item We systematically evaluate our approach on \changedFinalRevision{32}{41} real-world modeling projects and compare it against baselines that suggest changes (i) randomly, (ii) based on historical co-change frequency, and (iii) based on semantic similarity.
   \item We analyze factors contributing to \nameapproach predictive performance, including project size, multi-location change pattern size, and dispersion\changedFinalRevision{,}{ and} pattern characteristics \changedFinalRevision{and the effect of available historical data (cross-project setting)}{}.
   \item\changedFinalRevision{We evaluate \nameapproach in an iterative, multi-location completion setting by combining it with single-location model completion~\cite{tinnes2024leveraging} and compare it to single-location completion, performed $N$ times for next focus node prediction.}{}
\end{itemize}
\vskip -1ex
The dataset as well as the source code for \nameapproach{}, and the experiments are provided in \appendixsite{}.

%% file: Content/2_relatedwork.tex
\section{Related Work}


In this section, we provide an overview of existing work on model completion and the relation to other modeling activities.

\subsection{Model Completion (with LLMs) }

Previous work explores recommending model completions using pattern catalogs, where partial models are completed by identifying matching changes through pattern or graph matching and then applying the missing parts accordingly~\cite{kuschke2017rapmod, kuschke2013recommending, maeder2021, li2013efficient, deng2016recommendation, kogel2016automatic,kogel2017recommender}.

These approaches typically rely on domain-specific pattern catalogs that must be manually created and maintained. As a result, they are tied to a specific domain and modeling language, requiring new catalogs to be created for each domain-specific context, and they struggle with the verbose and noisy textual components found in software models. While rule-based approaches are explicitly defined and typically complete, this completeness can become a limitation when facing complex or underspecified scenarios, such as those encountered in model completion tasks.

As a consequence, efforts moving beyond rigid rule-based systems have been made. While these are more generalizable to broader applications, they focus on single-location model completion.

Initial steps from a natural language perspective have been taken by Agt-Rickauer et al.~\cite{agt2018domore, agt2019automated}, who use conceptual knowledge bases and semantic networks built from natural language data to suggest entity names of model elements. \citet{lopez2023word} train a skip-gram model to generate word embeddings specific to the modeling domain. They evaluate performance on meta-model classification, clustering, and an entity name recommendation task. 
Elkamel et al.~\cite{elkamel2016uml} recommend UML classes using clustering over existing model repositories, based on word-level similarities in names, attributes, and operations.
Burgue{\~n}o et al.~\cite{burgueno2021nlp} propose word embedding similarity to recommend domain concepts.

More recently, deep-learning models have been adapted to modeling tasks\changedFinalRevision{~\cite{di2025use}}{}. For example, Di Rocco et al.~\cite{di2022finding} use an encoder--decoder network to suggest element types to add in change-based persistence (CBP) models. As CBP is less common in practice~\cite{yohannis2020change}, we focus on state-based modeling instead. Weyssow et al.~\cite{weyssow2022recommending} trained a transformer-based model from scratch to suggest meta-model concepts, however, the effectiveness of such approaches is constrained by the limited availability of modeling data~\cite{chaaben2024utility}.
ModelMate~\cite{costa2024modelmate} is a recommender system designed for textual DSLs based on fine-tuned language models. The approach has been evaluated on a modeling task (predicting \texttt{EStructural\-Feature} names in Ecore meta-models) and compared against existing recommender systems~\cite{di2023memorec, burgueno2021nlp, weyssow2022recommending}. \citet{liu2024reco} propose an approach for predicting connections between modeling elements.

Chaaben et al.~\cite{chaaben2023towards, chaaben2024utility} use the few-shot capabilities of GPT-3 to suggest new model class names, attributes, and associations by providing examples from unrelated domains. The approach does not scale well to larger models, as it requires multiple queries depending on the model size and includes all model concepts in each prompt. 

Tinnes et al.~\cite{tinnes2024leveraging} concentrate on the neighborhood of the most recently changed element, thereby addressing prompt size limitations by restricting the scope of the LLM to a localized area. Their method was shown to outperform the approach by Chaaben et al.~\cite{chaaben2023towards} on industrial, real-world data. In addition, they incorporate domain-specific context through similarity-based few-shot retrieval from the software model repository.

In general, while various approaches for model completion have been proposed\changedFinalRevision{~\cite{di2025use}}{}, the focus has been on \textit{single-location changes}, with little attention to patterns that span multiple locations, possibly cutting across the entire software model.

\changedFinalRevision{\subsection{Supporting Other Modeling Activities}}{}

\changedFinalRevision{
Another line of research}{A further time of research } also uses models but does not consider model completion.
For example, a related area concerns ChatGPT’s model generation capabilities, either from natural language descriptions~\cite{camara2023assessment}, requirements~\cite{eisenreich2024requirements}, or images of UML class diagrams~\cite{conrardy2024image}. 
\citet{lopez2024text2vql} introduce a framework for generating model queries from natural language by fine-tuning open-source LLMs on a synthetic dataset created with ChatGPT. 
Other approaches provide similar examples of models through collaborative filtering~\cite{di2023memorec} and similarity-based filtering~\cite{di2023morgan}, but ultimately rely on users to apply the final model completion based on the examples~\cite{almonte2024engineering}. In the same vein, there is work on change impact analysis and trace link generation between different models, model types, and corresponding requirements artifacts, documentation, and code~\cite{awadid2023supporting,mackenzie2020change,anwer2024becia,fuchss2025lissa}.

Finally, there is the research area of meta-model co-evolution, where changes to the meta-model must be propagated to models and model transformations to maintain consistency~\cite{herrmannsdoerfer2008automatability, cicchetti2008meta}. These approaches aim to ensure correctness according to meta-model constraints and synchronization across modeling artifacts.
\changedFinalRevision{
Closely related to meta-model co-evolution is model repair~\cite{barriga2022ai, macedo2016feature}, which focuses on automatically correcting inconsistencies in software models.
Most approaches rely on graph transformation rules or OCL-like constraints to automatically repair models \cite{lauer2023empowering, ohrndorf2021b, nassar2017rule, mens2006detecting}, while others organize fixes (additionally) into repair trees~\cite{reder2012computing, marchezan2023generating}.}{}
\changedFinalRevision{
From a machine learning perspective, some model repair approaches use reinforcement learning, where rewards are based on achieving consistency and improving model quality~\cite{barriga2022parmorel, iovino2020model, barriga2019personalized}.}{}
\changedFinalRevision{In model completion, an oracle for checking consistency, is not available.}{}
In contrast \changedFinalRevision{to model repair and meta-model co evolution}{}, we additionally do not focus on meta-model conformance, but instead on maintaining and extending the semantic and functional aspects of software models during software evolution.
\vspace{5pt}
\subsection{Code Completion and Repair}

Challenges similar to multi-location model completion have been explored for source code. Many code-centered approaches enhance single-location code completion by incorporating repository-level context into LLM prompts via static analysis~\cite{ouyang2024repograph, liu2024stall, phan2024repohyper}. 

Regarding code co-changes and change impact analysis, considerable work has been done in recent years~\cite{li2013survey, zhou2024revealing, hong2024don, jaafar2011exploratory}.
\changedFinalRevision{}{For example, Zhou et al.~\cite{zhou2024revealing} represent changed code as a graph with spatial and temporal edges, then apply frequent subgraph mining to identify common change propagation channels. 
Hong et al.~\cite{hong2024don} use a graph neural network to recommend co-changed functions, modeling functions as nodes and co-change history as edges. They apply GraphSAGE to learn node embeddings and predict whether a function will be modified in a future commit.} 
For multi-location code completion, CodePlan \cite{bairi2024codeplan} converts a repository-level task into a plan graph of LLM-driven edit obligations discovered through incremental dependency and change-impact analyzes. It applies edits, recomputes affected dependencies, and iteratively extends the plan until all obligations are discharged. The resulting repository is then checked by an oracle; any failures become new input for the next cycle.
A related but distinct area focuses on LLM-based code repair~\cite{yang2024multi, wei2023copiloting,ye2024iter}, where models iteratively refine code using feedback from an oracle~\cite{ye2024iter}.

It is important to note that the approaches that work for code are not (easily) transferable to software models.
Unlike source code, software models are mostly non-executable artifacts that combine graphical structures with verbose textual annotations. This makes oracle-driven processes infeasible because candidate correctness cannot be validated automatically (e.g., via tests). In general, the field also suffers from a lack of publicly available datasets\changedFinalRevision{, which}{} significantly hinders comprehensive comparisons between different approaches~\cite{burgueno2025automation, tinnes2024leveraging, lopez2022modelset, muttillo2024towards}. Unlike source code, software models lack standard languages, formats, and evaluation metrics~\cite{izadi2022evaluation}, making benchmarking difficult. In contrast, code completion benefits from many benchmarks~\cite{wang2025software, paul2024benchmarks} like HumanEval~\cite{chen2021codex}.

%% file: Content/3_background.tex
\section{Preliminaries}

\subsection{Software Model Completion}

We represent software models as graphs to establish a common ground across different formats and types of software models, as is common in the literature~\cite{tinnes2024leveraging, 
lopez2022machine,
kchaou2017uml,
tinnes2023mining}.

\begin{definition}[Abstract syntax graph]\label{def:abstractsyntaxgrpah}
    An abstract syntax graph $G_m$ of a software model $m$ is an attributed graph, typed over an attributed type graph $\mathit{TG}$ given by metamodel $\mathit{TM}$.
\end{definition}

An attributed type graph $\mathit{TG}$ specifies the typing for abstract syntax graphs, ensuring that all elements conform to the structural and semantic constraints specified by the metamodel $\mathit{TM}$. For our purpose, we use a simplified representation of abstract syntax graphs as labeled directed graphs, where node and edge labels correspond to the textual names of their respective types and relations in the abstract syntax graph.

\begin{definition}[Labeled directed graph]\label{def:labeled_graph}
    A labeled directed graph $G$ over a label alphabet $L$ is defined as the tuple $(V, E, l)$, where $V$ is a finite set of nodes, $E \subseteq V \times V$ is the set of directed edges, and $l: V \cup E \to L$ assigns labels to nodes and edges~\cite{tinnes2024leveraging}.
\end{definition}

In a directed graph $G$, direct successors of a node \( v \in V \) are all nodes that are \emph{directly} reachable from \( v \) via an outgoing edge.

\begin{definition}[Direct successor set]

The direct successor set of a node \( v \) is defined as:
\vspace{-0.3em}
\begin{equation}
\text{succ}(v) = \{\ u \in V \mid (v, u) \in E \,\},
\end{equation}
where \( (v, u) \in E \) denotes a directed edge from \( v \) to \( u \).
    
\end{definition}

We further define the software model difference between successive software model versions.

\begin{definition}[Structural model difference]
    A \emph{structural model difference} $\Delta_{mn}$ of model versions $m$ and $n$ is obtained by matching corresponding model elements in the model graphs $G_{m}$ and $G_{n}$.
\end{definition}

The structural model difference $\Delta_{mn}$ contains changed elements $\Delta^\circlearrowright_{mn} = ((V_n \cup E_n) \setminus (V_m \cup E_m)) \cup ((V_m \cup E_m) \setminus (V_n \cup E_n))$ and preserved elements $\Delta^{=}_{mn} = (V_m \cup E_m) \cap (V_n \cup E_n)$\footnote{For simplicity, we omit the explicit matching and assume that $V_m$ and $E_m$ are identified with their matched counterparts, where applicable.}.

Next, we introduce local and multi-location model completions. 

\begin{definition}[Model completion]
A software model completion $\gamma_{(c,s)}$ transforms a given source model $m$, represented by $G_m$, into a (partial) target model $n$, represented by $G_n$~\cite{tinnes2024leveraging}: 
\vspace{-0.3em}
\begin{equation}
m \stackrel{\gamma_{(c,s)}}{\to} n,
\end{equation}
such that $\gamma$ corresponds to the model difference $\Delta_{mn}$. Here, $c$ denotes the number of elements involved in the completion change, that is, $c = \left| \Delta^\circlearrowright_{mn}\right|$ and $s$ is the maximum shortest-path distance between any pair of involved elements.
\label{multilocationdef}
\end{definition}

The parameter $s$ gives an indication of how spread the involved elements of a software model completion pattern are across the model. A small $s$ suggests that the change or pattern is locally confined, whereas a larger $s$ implies that the completion affects distant parts of the model. Therefore, we can define single-location changes and multi-location changes as follows:

\begin{definition}[Single-location model completion]
   A single-location software model completion is a model completion $\gamma_{(c,s)}$, where $s\leq1$.
\end{definition}

\begin{definition}[Multi-location model completion]
   A multi-location software model completion is a model completion $\gamma_{(c,s)}$, where $c>1$ and $s>1$.\footnote{
   Note that previous work~\cite{tinnes2024leveraging} has focused on
   single-location model completion with $c \leq 2$ and $s \leq 1$; That is, at most one new node and one connection to an existing element are added.}
\end{definition}

\begin{figure}[htbp]
  \centering
  \includegraphics[width=0.8\linewidth]{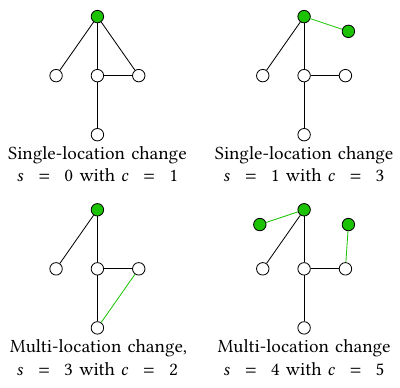}
\caption{Examples of single-location and multi-location software model changes with different values of $c$ and $s$, green element mark newly added elements}
  \label{fig:change_examples}
\end{figure}

 Examples for multi-location model completion and single-location model completion with different $s$ and $c$ are given in Figure~\ref{fig:change_examples}.

\subsection{Machine Learning}

Next, we outline the basic machine learning concepts required to understand our approach.

A \emph{feedforward neural network} defines a mapping 
\vspace{-0.3em}
\begin{equation}
y = f(x; \theta),
\end{equation}
where  \( x \) is the input and vector \( \theta \) contains the learnable parameters. The parameters are learned by minimizing the difference between predicted and target values during training~\cite{Goodfellow}.
Neural networks are composed of layers, where each layer applies a transformation to its input before passing it to the next layer, forming a hierarchical representation of the data. In classification, they map an input $x$ to an output category $y$. 
The learning process is guided by a loss function, which quantifies the error between the network's predictions and the actual target values~\cite{Goodfellow}.

An \emph{embedding model} is a representation learning model that transforms natural language data or other structured data into a lower-dimensional continuous vector space and is defined as:
\vspace{-0.3em}
\begin{equation}
\phi: X \to \mathbb{R}^d,
\label{embeddingformular}
\end{equation}
where \( X \) is the input space (e.g., words, graph nodes, model elements),  
\( \mathbb{R}^d \) is the \( d \)-dimensional vector space,  
and \( \phi(x) \) is the embedding of \( x \), capturing a subset of its properties in the lower-dimensional space.

%% file: Content/5_approach.tex
\section{Approach}\label{approach}

This section is structured as follows: we first define the general concept of multi-location model completion, then provide an overview of the \nameapproach{}'s workflow, followed by detailed descriptions of the data preparation process and \nameapproach{}'s different phases.

\subsection{Concepts}

\begin{figure}[htbp]
  \centering
  \includegraphics[width=\linewidth]{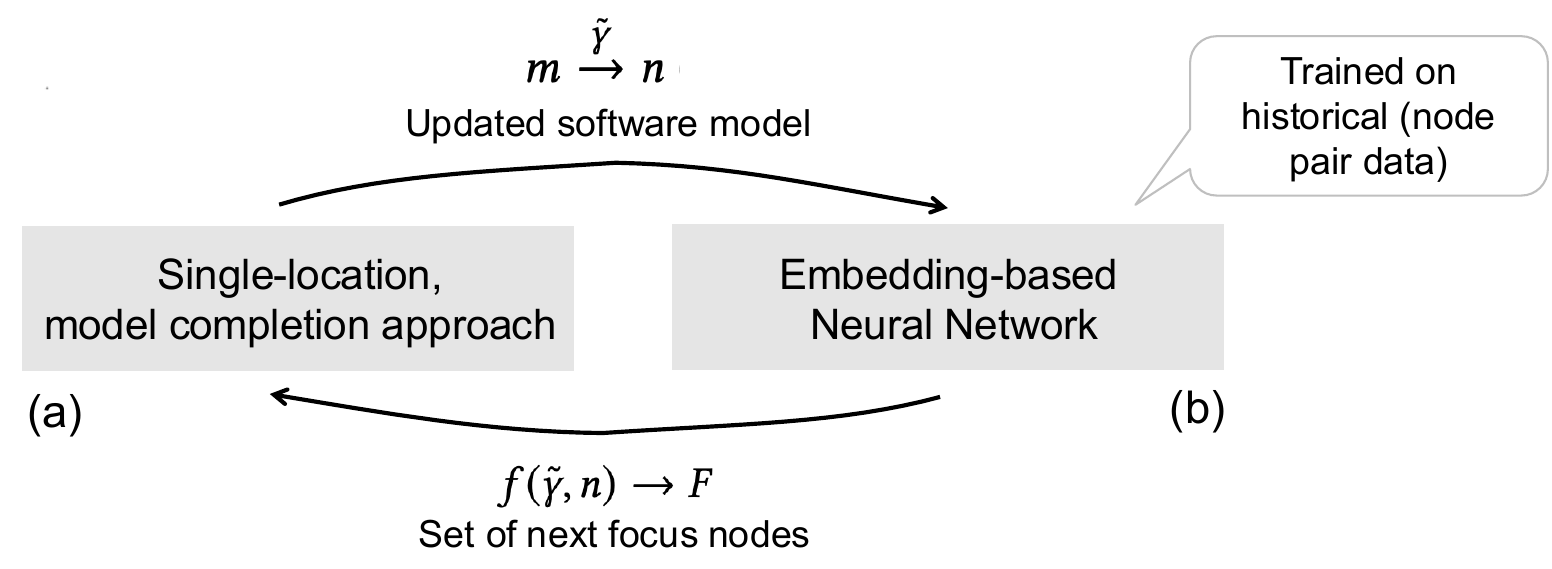}
\caption{Combined process of single-location model completion and next focus node prediction.}
  \label{fig:overview}
\end{figure}

Since LLMs combined with retrieval-augmented generation have already demonstrated strong performance for model completion tasks, even on real-world industrial data~\cite{tinnes2024leveraging}, we decompose the problem of multi-location changes into an iterative approach between single-location model completion and finding the next focus nodes, as illustrated in Figure \ref{fig:overview}. 
More specifically, a single-location, LLM-based model completion approach relying on an LLM (Figure \ref{fig:overview}, (a)) starts with a given software model~$m$, a slicing criterion~$\varsigma$, and a set of relevant elements~$C_m \subseteq V_m \cup E_m$\footnote{\citet{tinnes2024leveraging} use recently changed elements, where \citet{chaaben2023towards} use a small number of related classes(nodes)}. A slicing criterion~$\varsigma(C_m,m) \to C'_m$, is applied to extract the elements for the LLM context.\footnote{For example,~\citet{tinnes2024leveraging} use so called simple change graphs as a slicing criteria, but other options are also possible~\cite{chaaben2023towards}} These elements serve as the context for the LLM that performs a single-location software model completion: $m \xrightarrow{\mathclap{\tilde{\gamma}}} n$ (additional steps may be required depending on the approach~\cite{chaaben2023towards, tinnes2024leveraging, chaaben2024utility}).
As a next step, the global set of next focus nodes~$F$ needs to be predicted, enabling the overall approach to perform multi-location model completion (Figure \ref{fig:overview}, (b)), which is what we will explain next. First, we define the concept of a \textit{focus function}.
\begin{definition}

Given a source model $n$ and partial model completion $\tilde{\gamma} \subset \gamma_c$, with $c>1$, the \textit{focus function} is 
\begin{equation}
f(\tilde{\gamma}, n) \to F \subseteq V_n
\end{equation}
with the set of \textit{focus nodes} $F$ in $V$ of $n$. 
\end{definition}

The \textit{focus nodes} $F$ are then given back again to the slicing criteria~$\varsigma(C_n:=F,G_n)$ for single-location model completion in a new iteration.

\subsection{Workflow}

\begin{figure*}[h]
  \centering
  \begin{minipage}[t]{0.75\linewidth}
    \centering
    \includegraphics[width=0.9\linewidth]{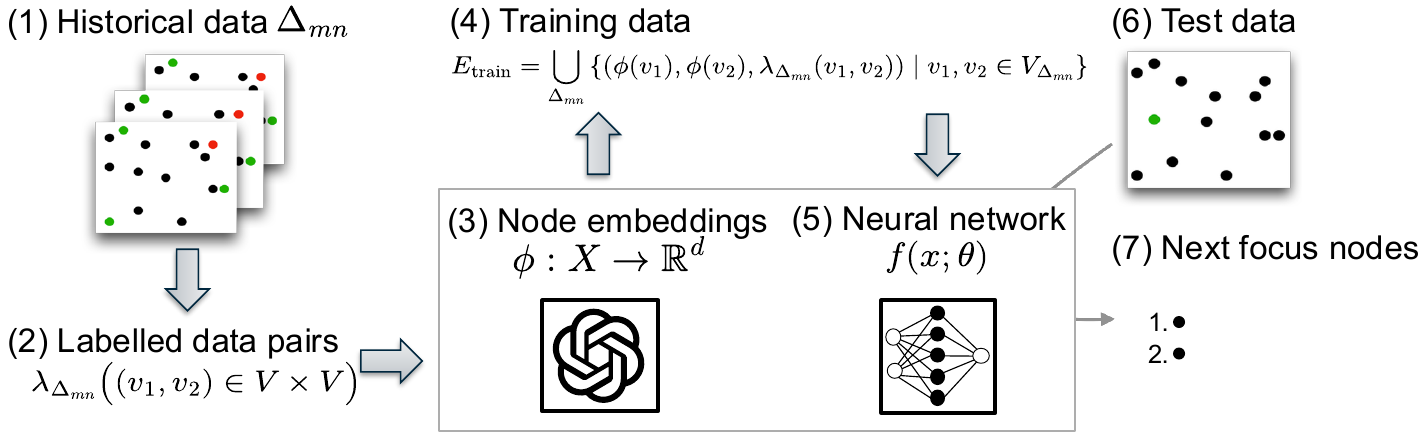}
      \captionsetup{font=small, skip=4pt}
    \caption{Overview of the \nameapproachfull\ (\nameapproach) approach.}
    \label{fig:overviewsteps}
  \end{minipage}%
  \hfill
  \begin{minipage}[t]{0.25\linewidth}
    \centering
    \includegraphics[width=0.7\linewidth]{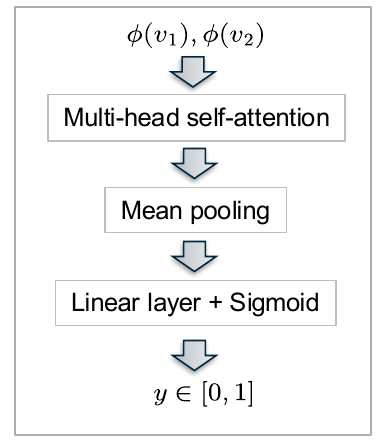}
      \captionsetup{font=small, skip=4pt}
\caption{Neural Network architecture of \nameapproach.}
    \label{fig:architecture}
  \end{minipage}
\end{figure*}

An overview of the workflow of \nameapproach is given in Figure \ref{fig:overviewsteps}. 
\nameapproach rests on a neural network that learns from historical data which elements tend to change together. In the first step, the training data is constructed (Figure \ref{fig:overviewsteps}, Step 1--4), including that each software model's nodes are embedded using an embedding model (Figure \ref{fig:overviewsteps}, Step 3). 
Node pairs are then passed through a neural network (Figure \ref{fig:overviewsteps}, Step 5) for training. For inference (Figure \ref{fig:overviewsteps},~Step~6), the software model's nodes are put though the embedding model and the neural network to evaluate their probability of changing together. Afterwards the probabilities are ranked, and the nodes with the highest scores are suggested as the next focus nodes (Figure \ref{fig:overviewsteps}, Step 7). In what follows, we describe the key phases in detail. 

\subsection{Data Preparation}

Following Tinnes et al.~\cite{tinnes2024leveraging}, we employ a model matcher -- specifically EMFCompare~\cite{brun2008model} -- to obtain the \emph{structural model differences} $\Delta_{mn}$ between each pair of consecutive models\footnote{In what follows, we discarded about $0.02\%$ of nodes due to their non-parsability.} (Figure \ref{fig:overviewsteps}, Step 1). These differences highlight the elements that have changed as well as those that have been preserved.
\changedFinalRevision{
Our approach operates on a graph-based representation of models and is thus not limited to Ecore~\cite{tinnes2024leveraging}.
The extraction can always be applied to a model difference (i.e., as long as model matching and differencing can be performed), and is therefore adaptable to a wide range of modeling tools. Often, model elements carry unique identifiers, making matching straightforward.
We transform the Ecore models into graphs with \texttt{networkx}, iterating over the detected changes to add them as nodes and connect them via their corresponding edges. This yields a general, notation-independent representation. The resulting}{Then, we prepare the} historical change data, which consists of sequential model versions \changedFinalRevision{is then prepared: t}{.T}he dataset is split into training, validation, and test sets, where the first is used as historical context for training, and the last is used for testing. Details on the specific splitting are provided in Section~\ref{evaluation}.

Given a model difference, we construct a set of node pairs and label those that have been modified in the same commit as positive examples (Label 1). These labeled pairs serve as ground truth for both neural network training and evaluation (Figure \ref{fig:overviewsteps}, Step 2). Unchanged node pairs are labeled as 0:
\vspace{-0.3em}
\begin{equation}
\lambda_{\Delta_{mn}}\bigl((v_1, v_2) \in V \times V\bigr) =
\begin{cases} 
    1, & \text{iff both } v_1 \ \text{and a direct }  \\
       & \text{successor of} \ v_2  \in \Delta^\circlearrowright_{mn} \\
    0, & \text{otherwise}.
\end{cases}
\label{eq:successorlabelassignement}
\end{equation}

\noindent

For clarity, we refer to recently changed elements in \( \tilde{\gamma} \), which were, for example, suggested by a single-location approach, as \emph{anchor nodes}. So, in Equation~\ref{eq:successorlabelassignement}, \( v_1 \) is the anchor node.

\subsection{Training Phase}

Given the training set consisting of model differences~$\Delta_{mn}$, we first apply a pre-processing step to balance the number of data points per software model. Specifically, we ensure that each model contributes an equal amount of training data, preventing the network from being biased towards larger software models with more data points. Then we embed each $v \in V_{\Delta_{mn}}$ according to Equation~\ref{embeddingformular} (Figure \ref{fig:overviewsteps}, Step 3). For this purpose, we have explored various embedding models, aiming to balance computational efficiency with the ability to capture essential differences in the data. After evaluating different options in a pilot study, we selected "text-embedding-3-small" with an embedding size of $e=1536$ from the OpenAI family.
Then, we input the embedded node representations pairs with their respective ground truth value into the neural network (Figure \ref{fig:overviewsteps}, Step 4--5). We use the Adam Optimizer for training.

\paragraph{Neural Network Architecture.}

Regarding the neural network architecture, we have explored linear and non-linear networks, but ultimately decided for an attention-based model~\cite{vaswani2017attention} \changedFinalRevision{. The overall architecture of our neural network is shown in Figure \ref{fig:architecture}. The embedded node representation pairs, given as the input, are first processed by a single self-attention layer,}{that performs a single round of self-attention} followed by mean pooling. 
A final linear layer maps the pooled representation to a single logit value per sample. At inference time, the logit value is passed through a sigmoid function to obtain a probability, while during training, the raw logit values are used directly with the loss function, which applies the sigmoid function internally.

\paragraph{Loss function.}

The neural network is trained using the binary cross-entropy loss (BCE), $l_{i}$
, which combines a sigmoid layer and the BCE loss for improved numerical stability. Given a minibatch \( \{(z_i,\,y_i)\}_{i=1}^N \), where 
\( z_i \in\mathbb{R} \) is the raw model output, 
\( y_i \in \{0, 1\} \) is the ground-truth label, and 
\( \hat{y}_i \;=\;\frac{1}{1+e^{-z_i}} \) is the predicted probability. 
\begin{equation}
\begin{aligned}
l_{i} =
\max(0, z_i) - z_i y_i + \log\left(1 + e^{-\left| z_i \right|}\right)
\end{aligned}
\end{equation}

To address extreme class imbalance, we apply a focal loss correction on top of the BCE formulation~\cite{lin2017focal}. This imbalance arises from the sheer number of negative examples (i.e., nodes that do not change together), which are often well-classified and would otherwise dominate the total loss.
We also add the focal loss weight to the loss term, which reshapes the loss function to down-weight easy examples.
\vspace{-0.3em}
\begin{equation}
w_{i} = 1 -( \hat{y}_i \cdot y_i + (1 - \hat{y}_i) \cdot (1 - y_i) )^\beta,
\end{equation}
\noindent
where \( \beta \) controls up-weighting of misclassified individual data points. As a result, false negative examples -- which may have been assigned high probabilities and are harder to classify using the standard BCE loss -- contribute more to the training process, effectively pushing them out of the set of predictions with the highest probabilities, which will be later important for ranking.

While $w$ focuses on individual data points, we additionally apply a class-level balancing factor $a$.
\vspace{-0.3em}
\begin{equation}
a_{i} = \alpha \cdot y_i + (1 - \alpha) \cdot (1 - y_i)
\end{equation}

Additionally to the focal loss terms \changedFinalRevision{introduced by Lin et al.~\cite{lin2017focal},}{~\cite{lin2017focal},} to optimize for our recommendation task, we add a misclassification penalty for false negatives.
\begin{equation}
m_{i} = (1 - y_i) \cdot  \hat{y_i }\cdot \lambda + 1,
\end{equation}
\noindent
where \( \lambda \) is the penalty scaling factor for incorrect high-probability negatives.
Combining these components, the final focal loss function for our task of predicting next focus nodes is:
\vspace{-0.3em}
\begin{equation}
\mathcal{L} = \frac{1}{N} \sum_{i=1}^{N}  \alpha_i \cdot w_i \cdot l_i \cdot m_i
\end{equation}
\vspace{-0.3em}

\subsection{Inference Phase}\label{inferencephase}

Given a model represented by a labeled directed graph \( G \) and a partial model completion \( \tilde{\gamma} \), for example, obtained from a single-location model completion approach (Figure \ref{fig:overviewsteps}, Step 6), \nameapproach suggests the next~\textit{focus nodes} in the inference phase (Figure \ref{fig:overviewsteps}, Step 7). The anchor node \( v_1 \in \tilde{\gamma} \) that has been changed is embedded, and \nameapproach computes the probability of each other node \( v \in V \), where \( v \neq v_1 \), changing together with \( v_1 \). According to Equation~\ref{eq:successorlabelassignement}, a change is expected to occur at a (direct or indirect) successor of \(v\), either through addition, deletion, or modification, which then can be suggested by a single-location model completion approach.

The node pairs \( (v_1, v) \) with  \( v \in V \) are passed to our trained neural network, which computes the probability \( f(\phi(v_1), \phi(v)) \) based on the historical evolution of the current software model and patterns learned from other models.

Finally, we rank all nodes \( v \neq v_1 \) based on the probability of changing together with \( v_1 \) and suggest the top-\(k\) candidates as the next focus nodes, which can then be presented to the user or be fed into the next iteration (Figure \ref{fig:overview}, Step (a)).

%% file: Content/6_experiments.tex
\section{Evaluation}\label{evaluation}

We empirically evaluated \nameapproach using historical real-world modeling data to assess its ability for multi-location model completion. 
Working with historical data allows us to separate the technical capabilities of our approach from other factors introduced by tools, such as the optimal number of recommendations shown, the layout and positioning of modeling elements, or tool-specific evaluation metrics. This facilitates a reproducible and comparable assessment of the capabilities of our approach.

In what follows, we outline our research questions, describe the evaluation setup and data used, and present our results.

\subsection{Research Questions}

We are interested in whether our \nameapproach, given a change that has been applied (i.e., an anchor node), can effectively predict the next focus node(s) in a multi-location model completion task. Specifically, we examine whether a model trained on historical multi-location changes is able to generalize to new, unseen changes.

\begin{tcolorbox}[arc=0pt, outer arc=0pt, boxrule=0.5pt, top=1pt, bottom=1pt, left=2pt, right=2pt, breakable,  sharpish corners,enhanced]
\noindent\textbf{RQ~1:} \textit{To what extent can \nameapproach predict new focus nodes for multi-location software model completion?}
\end{tcolorbox}

To better understand the \nameapproach predictive performance, we investigate how its performance varies with the distance between the predicted focus node(s) and the originally changed (anchor) node, 
that is, how the performance of the model completion \( \gamma_{(c,s)} \) depends on \( s \).
In particular, we examine whether the model is better at predicting single-location changes (close in terms of graph distance) or also performs well on more global changes.

\begin{tcolorbox}[arc=0pt, outer arc=0pt, boxrule=0.5pt, top=1pt, bottom=1pt, left=2pt, right=2pt, breakable,  sharpish corners,enhanced]
\noindent\textbf{RQ~2:} \textit{How does the model’s predictive performance of new focus nodes depend on the distance (in terms of graph radius) to the anchor node? }
\end{tcolorbox}

\changedFinalRevision{We also}{Finally, we } investigate the conditions under which our \nameapproach performs well and identify scenarios in which its predictive performance could be improved. Specifically, we examine which project specific properties influence the model's ability to correctly identify new focus nodes. These properties include, for example, the overall project size (i.e., the number of training data points), the proportion of positive instances (i.e., data points with a ground truth of one), and the kind and content of the change patterns.
\changedFinalRevision{On the other hand, we study the performance of \nameapproach in a cross-project setting, that is, whether \nameapproach generalizes to previously unseen projects by transferring known
project-specific characteristics. This aspect becomes particularly relevant in real-world scenarios, if no historical data are available for a given project.}{}

\begin{tcolorbox}[arc=0pt, outer arc=0pt, boxrule=0.5pt, top=1pt, bottom=1pt, left=2pt, right=2pt, breakable,  sharpish corners,enhanced]

\noindent\textbf{RQ~3:} \textit{Which project-specific or pattern-specific properties influence the predictive performance of our model \changedFinalRevision{and
consequently how well does \nameapproach perform in a cross-project setting}{}?}
\end{tcolorbox}

\changedFinalRevision{Finally, we investigate the performance of \nameapproach in a complete multi-location model completion setting (not only next focus node prediction) that is obtained by iteratively combining it with a single-location model completion approach, as shown in Figure~\ref{fig:overview}.}{}

\begin{tcolorbox}[arc=0pt, outer arc=0pt, boxrule=0.5pt, top=1pt, bottom=1pt, left=2pt, right=2pt, breakable, sharpish corners, enhanced]

\noindent \changedFinalRevision{\textbf{RQ~4:} \textit{How effectively does \nameapproach support iterative, multi-location model completion?}}{}

\end{tcolorbox}

\subsection{Experiment Setup}

We conducted \changedFinalRevision{six}{four} experiments to address the \changedFinalRevision{four}{three} research questions; Experiments 3 \changedFinalRevision{,4 and 5}{and 4 both} contribute to answering RQ3.

\paragraph{Data.}
For all experiments, we use a publicly available, real-world dataset, RepairVision~\cite{ohrndorf2021history,ohrndorf2021b}, which contains versioned modeling projects. This is essential for our study, as it provides us with ground-truth information on multi-location changes that have been \emph{actually} performed by modelers in a real-world scenario\footnote{Other datasets such as the ModelSet~\cite{lopez2022modelset} contain only static snapshots, which would require synthetically constructing modeling histories. This does not reflect a real-world scenario and introduces confounding assumptions.}.

In total, the data set contains 41 modeling projects, with 912 commits. On average, the models contain 1285.9 nodes, and there are 168.9 changes per commit. 
For our evaluation, we applied an additional filtering step (e.g., because we required projects to have, at least, three commits to allow for a valid train/validation/test split) resulting in 32 projects considered in total.
Detailed information on each project and filtering is provided in \appendixsite{}. We use EMFCompare’s model matching capabilities to compute structural model differences for all modeling projects.

\paragraph{Experiment 1.} \label{experiment1}

To answer RQ 1, we split the modeling dataset into training, validation, and test sets while respecting the historical timeline.
More specifically, given the historically ordered structural model differences $\{\Delta_{m_1m_2}, \Delta_{m_2m_3}, \dots, \Delta_{m_{n-1}m_{n}}\}$, where $n$ is the number of structural model differences in a project, we split by commit, i.e., by structural model difference, to prevent data leakage between sets. We define the training set as $\{\Delta_{m_1m_2}, \dots, \Delta_{m_{n-3}m_{n-2}}\}$ , the validation set as $\{\Delta_{m_{n-2}m_{n-1}}\}$, and the test set as $\{\Delta_{m_{n-1}m_{n}}\}$. Overall, this leads to a ratio of 71.88\% train, 16.41\% validation, and 11.70\% test data points in the respective sets.\footnote{ 
To ensure realistic evaluation, we approximate a commonly used data ratio for train–validation–test splits (around 70–80\% train, 10–15\% validation/test).
Since each structural model difference can contain a highly variable number of data points (node pairs, see Equation \ref{eq:successorlabelassignement}), especially, in later commits, where models tend to be larger, we had to restrict the number of structural model differences in the validation and test set. Otherwise, those sets would have ended up with more data points than the training set, despite covering fewer commits. On the other hand, the neural network is trained on individual data points rather than entire commits, which leads to the specific dataset split proportions used.
}
Using commit time for splitting, rather than random sampling, mirrors a real deployment: We train on what is known, the commit history, and expect the network to generalize to new, unknown software models in the test set. 

We begin by preprocessing the data
(see \Cref{approach}) and training the neural network on the training set while tuning hyperparameters on the validation set. During training, we explicitly over-sampled or under-sampled data points from each project to a fixed size, ensuring that the neural network treats each project equally rather than being biased toward larger datasets.

We tuned all hyperparameters using Bayesian optimization, more information is given in \appendixsite{}. The task is framed as a node-ranking problem: Given a recently changed (anchor) node, the model ranks other nodes based on the probability of changing with this anchor node.

For evaluation purposes, we take models from the test set, which include the latest changes in the modeling history $\{\Delta_{m_{n-1}m_{n}}\}$, specifically the transition from the second-to-last to the last model snapshot. Given a recently changed element in  
\( \Delta^\circlearrowright_{m_{n-1}m_{n}} \), the anchor node,
and the set of already existing elements  
\( \Delta^{=}_{m_{n-1}m_{n}}\),  
we predict the next focus node(s), that is, the element whose successor is expected to be changed next (see Equation \ref{eq:successorlabelassignement}). That is, for evaluation, we remove the ground truth elements \( \Delta^\circlearrowright_{m_{n-1}m_{n}} \) (changes that have been made by the modeler in a real-world scenario) from \( \Delta_{m_{n-1}m_{n}} \) and investigate whether \nameapproach is able to predict these correctly.
We are particularly interested in the overall predictive performance of our \nameapproach. Neural network performance is commonly evaluated using Precision@$k$ on the test set~\cite{roy2022systematic, mcelfresh2022generalizability,capuano2022learning, tamm2021quality}\footnote{We do not report recall, as the number of relevant items varies significantly across cases -- from over 1000 to as few as 1--2, making recall highly sensitive to the denominator and thus difficult to interpret. Instead, we focus on top-$k$ precision, which better aligns with our recommender system setting. The goal is to recommend the most likely next changes first -- not to recover all possible changes. We additionally include a random baseline for comparison. Including a baseline that selects candidates randomly provides a meaningful lower bound and allows for relative performance assessment without relying on absolute metrics like recall.}.


A prediction is considered correct if the suggested node(s) were indeed modified in the corresponding commit in the dataset.

Let \( y_i \in \{0, 1\} \) be the binary ground-truth label for node \( i \), where 1 indicates that $i$ changed, we define \(\text{Precision@}k\) as the number of true positives among the top-\(k\) predictions, normalized by the minimum of \(k\) and the number of actual positives:

\begin{equation}
\text{Precision@}k = 
\frac{\# \text{true positives in top-}k}{\min(k,\ \# \text{actual positives})}
\end{equation}

With regard to \( k \), the number of recommendations, prior work consistently suggests keeping recommendation lists short and manageable for human users.  
Therefore, we limit \( k \) to a maximum of 10, but we report results for various values of \( k \leq 10 \), as well~\cite{ kuschke2013recommending, adhikari2024simima, chen2025need}.

We compare \nameapproach against three baselines: (i) \Random{} of focus nodes, (ii) \Semantic{} based on pre-trained embeddings, and (iii) \Historical{}, which prioritizes nodes that have frequently changed together in the past. We selected these baselines to reflect fundamentally different strategies for focus node prediction: 
(i) \Random{} serves as a naive lower bound, illustrating how well the other approaches perform compared to uninformed guessing; 
(ii) \Semantic{} builds on the assumption that semantically related elements tend to co-change\changedFinalRevision{, a concept also used in related work~\cite{kagdi2013integrating, kchaou2017uml, agt2018domore, agt2019automated, lopez2023word, elkamel2016uml, burgueno2021nlp}
}{}, and 
(iii) \Historical{} builds on the assumption that elements which changed together in the past are likely to do so again\changedFinalRevision{~\cite{kagdi2013integrating, zimmermann2005mining, jiang2021investigating}}{}. Together, these baselines cover a broad range of factors that can influence performance.

For the \Semantic{} baseline, we use the same embedding model as the one described in \Cref{approach}. Given the anchor node, we compute the cosine similarity between its embedding and those of all other nodes in the software model. The top-$k$ most similar nodes are then recommended. While there is currently no multi-location model completion approach available that we could adopt as a baseline, we use \Semantic{} as a reference point due to its significance in related domains. For instance, text-based similarity 
has been applied for change impact analysis~\cite{kagdi2013integrating} on source code, and in the UML model domain~\cite{kchaou2017uml}. 
Prior efforts for single-location model completion also~\cite{agt2018domore, agt2019automated, lopez2023word, elkamel2016uml, burgueno2021nlp} focused on similarity.

For the \Historical{} baseline, we construct a co-change matrix that records how often each pair of nodes has changed together in past commits. During inference, we identify the top-$k$ nodes with the highest co-change frequency with respect to the given anchor node and recommend those. We are interested in the overall performance, so we examine the overall distribution of Precision@$k$ values.
Historical co-change frequency has been frequently used on source code~\cite{kagdi2013integrating, zimmermann2005mining, jiang2021investigating}.

\paragraph{Experiment 2.}

To investigate how \nameapproach performs on multi-location change patterns of varying size, we limit the predicted focus nodes to a certain radius. That is, we only consider \( \gamma_{(c, s)} \) with \( s < \tau \), where \( s \) is the maximum shortest-path distance between any pair of involved elements in the multi-location change (Definition \ref{multilocationdef}) and \( \tau \) is a radius threshold. This setup allows us to analyze whether the model performs better on localized changes.  
By increasing \( \tau \), we study whether \nameapproach maintains high \(\text{Precision@}k\) even as changes become more spread across the model. We train the neural network using the same setup as in Experiment 1.

\paragraph{Experiment 3.}

We investigate how specific project properties influence the overall performance of \nameapproach. As a first step, we examine \nameapproach's average performance across individual modeling projects. We also analyze the influence of the overall training set size per project and the number of positive examples included in each project's training set. While neural networks typically benefit from more data points seen during training, we aim to understand whether this correlates with higher average performance per project. Note that we explicitly over-sample or under-sample during training to normalize the number of data points per project. This ensures that \nameapproach treats each project equally and avoids biasing towards datasets with more training examples.
The neural network is trained as in Experiment 1.

\paragraph{Experiment 4.}

To answer RQ3, we manually analyze the graphs in our test set to examine which change patterns work well and which do not. We additionally summarize the change, determine whether the single-location changes truly belong together or occurred by coincidence, and identify the overall pattern. For additional support, we consulted \textsc{OpenAI’s GPT model (o3)}.
\changedFinalRevision{\paragraph{Experiment 5.}

Finally, to address the last aspect of RQ3, which is, whether \nameapproach generalizes to a cross-project setting, we split the training and test sets by project rather than by historical data within a single project. The setup follows the idea of 10-fold cross-validation, where in each fold one project is used for testing and the remaining projects are used for training. Further details of the setup are given in \appendixsite{}. This setup allows us to evaluate how well the model, trained on certain projects, generalizes to previously unseen projects. We additionally compare it against a setup where historical data is available (Experiment 3).}{}%
\changedFinalRevision{\paragraph{Experiment 6. \label{experiment6}}
To answer RQ4, we first evaluate the performance gain of \nameapproach by comparing it against a state of the art, single-location model completion approach. Specifically we compare \nameapproach to the approach by Tinnes et al., called \nametinnes ~\cite{tinnes2024leveraging}.

We assess how \nametinnes performs when used iteratively for \textit{next focus node prediction} on its own.
For baselining, we use the same LLM, slicing procedure, prompt structure, linearization format, which represents the graph solely by edges, and the same database for few-shot retrieval from~\citet{tinnes2024leveraging}. The only, but crucial difference lies in the procedure of next focus node prediction, which is either done by \nameapproach (our approach) or indirectly by the LLM itself
given via the source node of the suggested edge (approach of~\citet{tinnes2024leveraging}). We perform the model completion for each data point, $N$ times, set $N = 10$, and report the results for next focus node prediction. For \nameapproach{}, we fix $k = 1$ to make the approaches comparable in this iterative scenario.
Exact details on the methods are provided in \appendixsite{}.

Second, we combine \nameapproach with the single-location model completion, \nametinnes~\cite{tinnes2024leveraging} and further improve \nametinnes for the multi-location setting, which we call \nametinnesbetter. This step is necessary because \nametinnes is not designed for iterative, multi-location completion. In particular, its edge-only linearization restricts the ability to represent jumps to new or isolated nodes in the model. As a result, when the next focus node is not yet structurally connected to previously completed regions, \nametinnes cannot represent this new node, leading to a drop in performance in multi-location settings.

The \textit{multi-location model completion} is performed iteratively, interleaving the prediction of the next focus node (\nameapproach), performing local model completion (\nametinnes or \nametinnesbetter), and updating the model under construction based on the completion result (see Figure~\ref{fig:overview}). The updated model is then passed to the next iteration. We call the settings {\nameapproach}+\nametinnes and {\nameapproach}+\nametinnesbetter and compare it against \nametinnes.
\nametinnesbetter is directly based on \nametinnes, but replaces the slicing
procedure with radius-based slicing and switches the graph serialization from
\textsc{EdgeL} to \textsc{JSON}. The prompt is slightly adapted accordingly, and we use a newer chat-based
LLM version (no major performance impact and better forward comparability; see~\appendixsite{}). The underlying database remains unchanged. As evaluation metrics, we both report the correctness of the next focus node and the resulting single-location model completion. We distinguish three levels of correctness for the model completion according to \citet{tinnes2024leveraging}: \textit{format correctness}, \textit{structural correctness} (valid graph structure and connections), \textit{change structure correctness} (correct change types such as add, modify, or remove), and \textit{type structure correctness} (exact type and change type). Exact details are on~\appendixsite{}.
}{}

\subsection{Results}

\paragraph{Experiment 1.}

We first focus on the overall Precision@\(k\) of \nameapproach for multi-location software model completion, comparing it to the \Random{}, \Historical{}, and \Semantic{} baselines.  

\begin{figure}[htbp]
    \centering
    \includegraphics[width=\columnwidth]{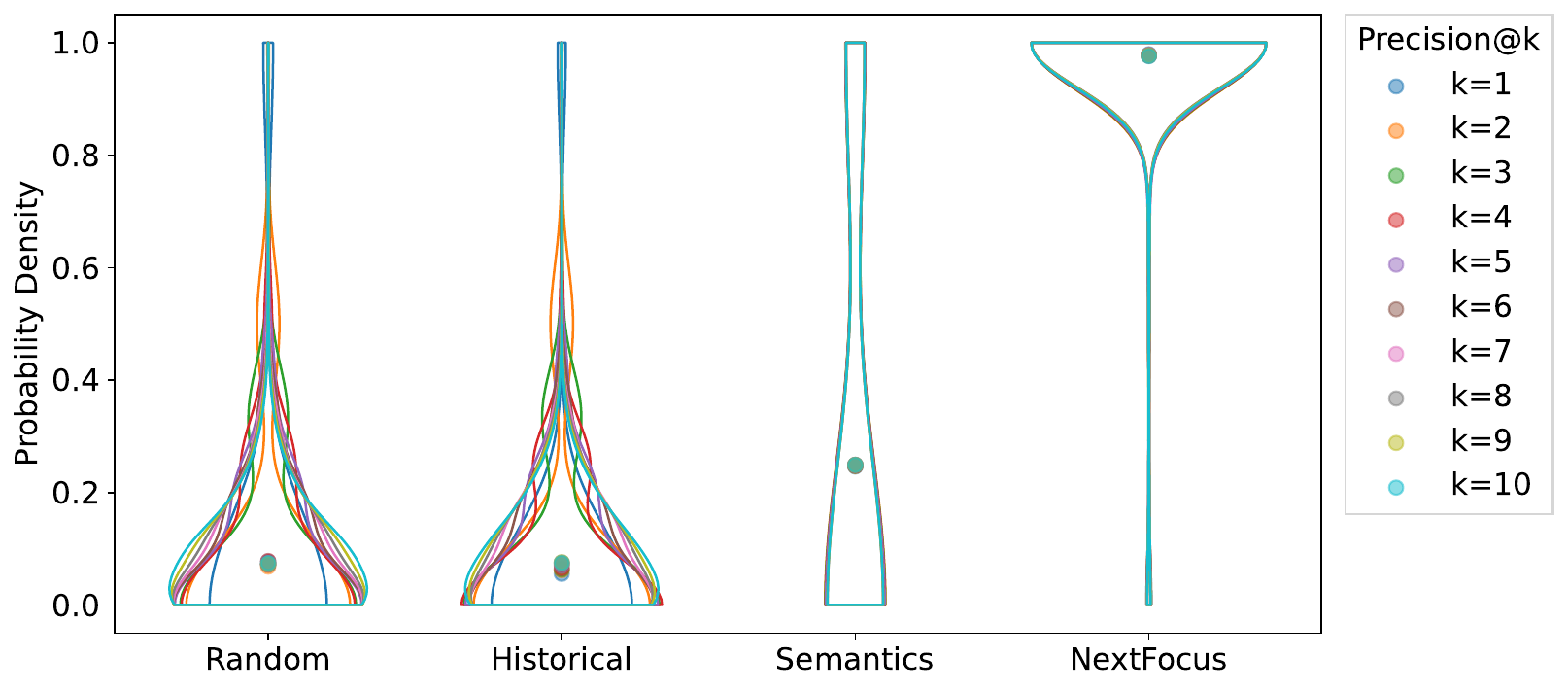}
    \captionsetup{font=small, skip=4pt}
    \caption{Precision@$k$ distribution of \Semantic{}, \Historical{}, \Random{}, and \nameapproach.
    }
    \label{fig:yallvalues}
\end{figure}

Figure~\ref{fig:yallvalues} presents a comparison of all approaches for values of \(k \leq 10\). We calculate the overall mean of the precision@$k$ values for each approach by averaging across all $k \in \{1, \dots, 10\}$. 
Overall, \nameapproach performs best, achieving an average of 0.98. It is followed by the \Semantic{} baseline (0.25), the \Historical{} baseline (0.07), and the \Random{} baseline (0.07).
We conducted one-sided Mann-Whitney U tests to assess statistical significance. \nameapproach significantly outperformed all baselines at every $k \in \{1,\dots,10\}$ ($p < 0.01$).
Among the baselines, \Semantic{} consistently outperformed both \Historical{} and \Random{} across all $k$ ($p < 0.01$). Exact $p$-values are provided in \appendixsite{}.

\begin{tcolorbox}[arc=0pt, outer arc=0pt, boxrule=0.5pt, top=1pt, bottom=1pt, left=2pt, right=2pt, breakable,  sharpish corners,enhanced]
\noindent \textbf{Summary Experiment 1:} \textit{\nameapproach significantly outperforms all baselines in terms of Precision@\(k\) (\(k \leq 10\)), with the highest average precision of 0.98.
}
\end{tcolorbox}

\paragraph{Experiment 2.}

\begin{figure}[h]
    \centering
    \includegraphics[width=\columnwidth]{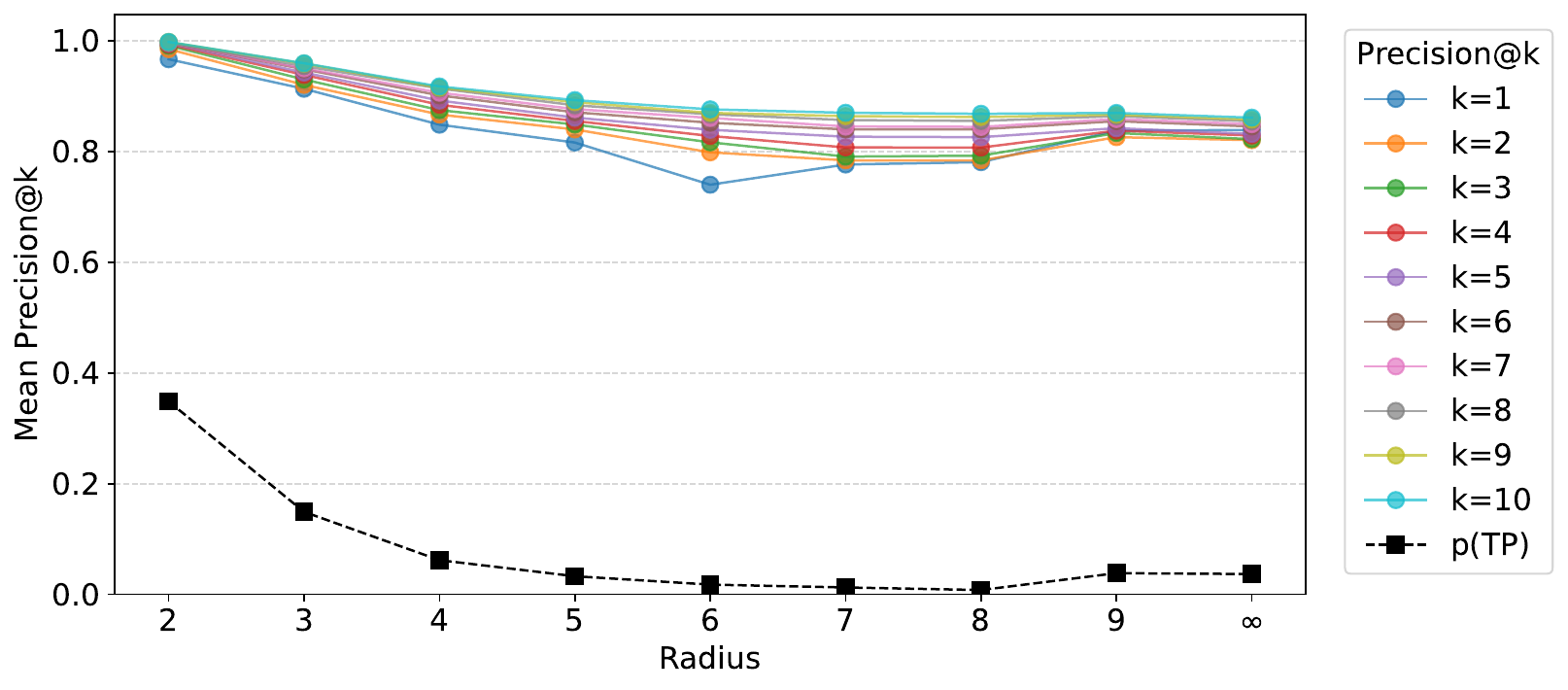}
\captionsetup{font=small, skip=4pt}
\caption{\nameapproach's performance with regard to the maximum radius considered}
    \label{fig:Figure_radius}
\end{figure}

In Figure~\ref{fig:Figure_radius}, we show the performance of \nameapproach depending on the considered radius. We limit the radius to the maximum values observed. Some of our graphs are disconnected, hence the value infinity for the distance ($s = \infty$).
We observe a generally negative monotonic relationship between radius and Precision@\( k \), with a Spearman's correlation coefficient \( \rho \) ranging from \(-0.19\) (Precision@3) to \(-0.14\) (Precision@1). This indicates that absolute Precision@$k$ slightly decreases with increasing radius.

We additionally plotted Precision@$k$ for random guessing (Figure~\ref{fig:Figure_radius}, p(TP)). For Precision@$k$, randomly selecting items yields an expectation equal to the overall prevalence of positives, independently of $k$. 
Since more nodes become candidates with increasing radius, making it harder for the model to identify relevant nodes, we additionally examined performance relative to random selection.
\nameapproach's performance improves relative to the \Random{} baseline, as indicated by a positive monotonic relationship between radius and the ratio of Precision@$k$ to the prevalence of positives. Spearman's correlation coefficients for this ratio range from \( \rho = 0.465 \) (Precision@1) to \( \rho = 0.625 \) (Precision@10) with $p<0.01$.
Using the additive margin over chance (\(\text{Precision}@k - p(\text{TP})\)), 
we again observe a positive monotonic relationship with the radius, 
Spearman’s \(\rho\) ranges from \(0.424\) (Precision@2) to \(0.515\) (Precision@10) (all \(p<0.01\)).

\begin{tcolorbox}[arc=0pt, outer arc=0pt, boxrule=0.5pt, top=1pt, bottom=1pt, left=2pt, right=2pt, breakable,  sharpish corners,enhanced]
\noindent \textbf{Summary Experiment 2:} \textit{We observe a slight negative monotonic relationship between maximum radius of the multi-location model completion and absolute Precision@\(k\), but a positive monotonic trend for the ratio of Precision@\(k\) to positive prevalence.
}
\end{tcolorbox}
\begin{figure}[bp]
    \centering
    \includegraphics[width=\columnwidth]{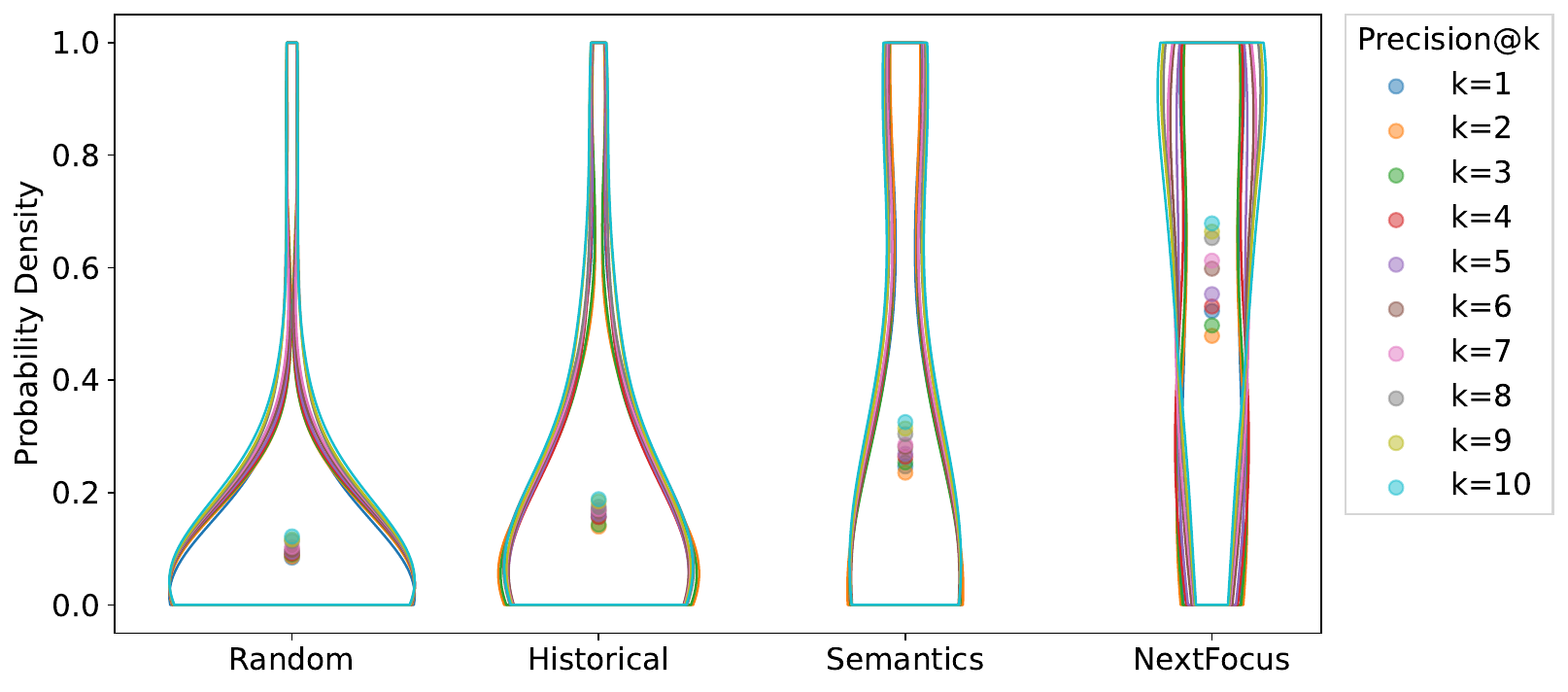}
    \captionsetup{font=small, skip=4pt}
    \caption{Distribution of average Precision@$k$ per project for the \Semantic{}, \Historical{}, \Random{} and \nameapproach approach.
    }
    \label{violinavgprojects}
\end{figure}
\paragraph{Experiment 3.}

We examine \nameapproach’s average performance across individual modeling projects, as shown in Figure~\ref{violinavgprojects}, which depicts the distribution of the predictive performance. While the overall performance remains higher for the \nameapproach (0.58) than for the baselines, individual project outcomes vary, with some projects performing notably better. A Kruskal--Wallis test confirms that these differences are statistically significant across all $k$ ($p<0.01$).

\begin{figure*}[t]
    \centering
    \begin{minipage}[t]{0.49\textwidth}
        \centering
        \includegraphics[width=\linewidth]{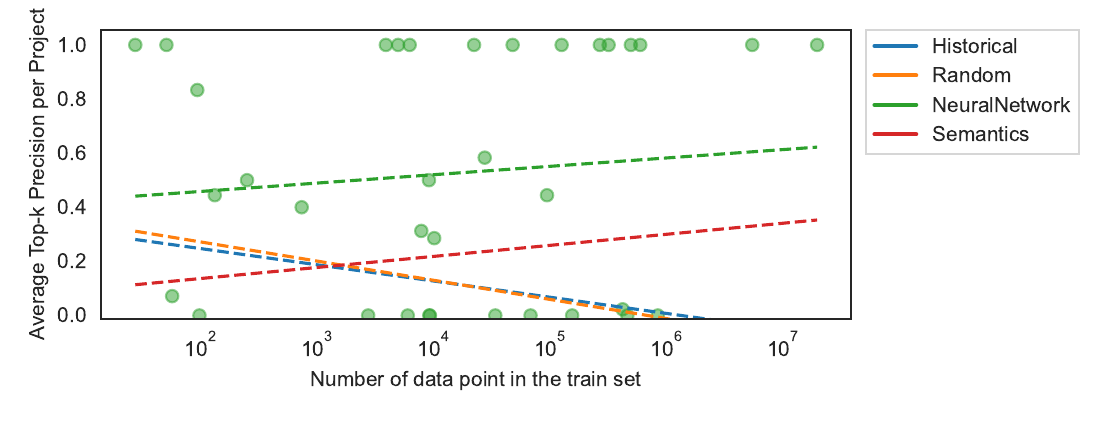}
        \captionsetup{font=small, skip=4pt}
        \caption{Average project Precision@$k$ over $k$ on the test set compared to the total number of training data points per project.}
        \label{fig:figtotalvalues}
    \end{minipage}
    \hfill
    \begin{minipage}[t]{0.49\textwidth}
        \centering
        \includegraphics[width=\linewidth]{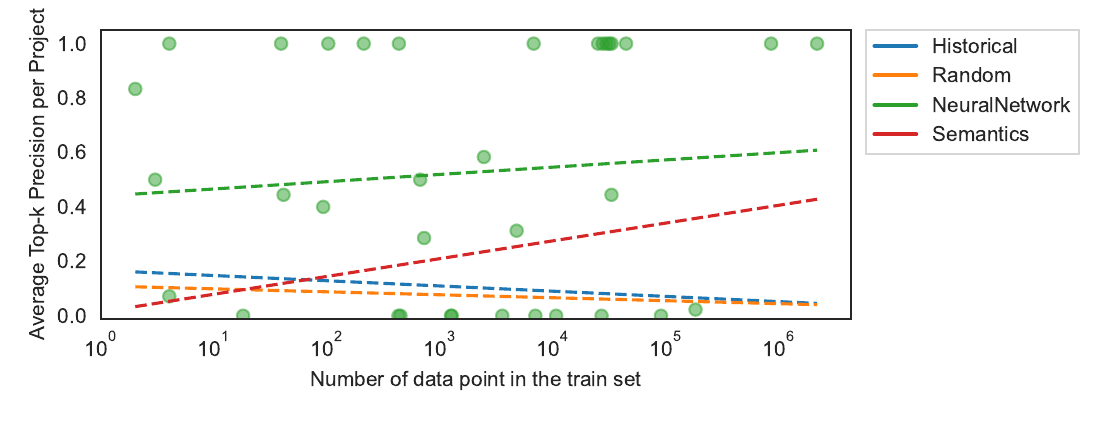}
        \captionsetup{font=small, skip=4pt}
        \caption{Average project Precision@$k$ over $k$ on the test set compared to the number of training positives (label=true) per project.}
        \label{fig:figtruevalues}
    \end{minipage}
\end{figure*}

We are particularly interested how the overall training set size and the number of positive examples in the training influence model performance. To visualize overall trends, we plot the relationship between dataset train size and the number of ground truth label equal to true and \nameapproach's average project Precision@$k$ over all $k \leq 10$, fitting a separate linear regression (Figure \ref{fig:figtotalvalues}, \ref{fig:figtruevalues}).
\changedFinalRevision{The green dots indicate the average predictive performance per project.}{}
To ensure a fair comparison across projects with varying candidate set sizes, we choose \(k\) dynamically as a small fraction of the total candidate count (e.g., \(k = \lceil 0.01 \cdot \text{candidates} \rceil\)).
We find no statistically significant monotonic relationship between dataset train size and average project Precision@$k$ for all approaches (Figure~\ref{fig:figtotalvalues}).

Analyzing the correlation between the number of positive examples and performance (Figure~\ref{fig:figtruevalues}), we find no significant trend for the \Historical{}, \nameapproach, and \Random{} ($p > 0.05$). Only \Semantic{} shows a statistically significant weak positive correlation ($\rho = 0.35$, $p = 0.040$) \cite{schober2018correlation}.

\begin{tcolorbox}[arc=0pt, outer arc=0pt, boxrule=0.5pt, top=1pt, bottom=1pt, left=2pt, right=2pt,  sharpish corners,enhanced]
\noindent \textbf{Summary Experiment 3:} \textit{Performance significantly varies between different projects, but \nameapproach still consistently outperforms the baselines across projects. No strong correlation is found between train dataset size or ground truth label count in the train dataset.
}

\end{tcolorbox}

\paragraph{Experiment 4.}

In contrast to the baselines, \nameapproach showed particularly strong performance in several scenarios, especially where common patterns were applied. Notable cases of high predictive performance included changes that involved renaming or replacing existing types, as well as identifier updates. 
One example of such a case was replacing the enum \texttt{ValidationSetType} with a new one \texttt{AggregationType}.

Some software model extensions also yielded strong results,  examples include additions of entirely new modeling concepts, such as the introduction of an \texttt{IfStatement} element to support if-then-else branching.
In total, there were three projects in which all approaches performed well due to the high probability of selecting the correct target among the candidates. 
Examples with low precision (below 0.2) include changes in the modeling hierarchy. 
\nameapproach did not perform well on all changes that introduced entirely new modeling concepts 
and performed worse on changes that shifted the underlying meaning of elements, such as adding new behavioral constructs or loosening attribute constraints, for example, the introduction of a \texttt{Trigger} concept for event-action logic and the removal of the uniqueness constraint 
from string-valued attributes.

\begin{tcolorbox}[arc=0pt, outer arc=0pt, boxrule=0.5pt, top=1pt, bottom=1pt, left=2pt, right=2pt,  sharpish corners,enhanced]
\noindent \textbf{Summary Experiment 4:} \textit{\nameapproach excelles in scenarios with recurring patterns but also performed well on some model extensions. It is less effective for hierarchy-related changes. \nameapproach outperformed the baseline approaches in almost all situations.
}

\end{tcolorbox}

\vspace{-0.85ex} 
\changedFinalRevision{\paragraph{Experiment 5.} Next, we analyse \nameapproach’s average performance in a cross-project setting. Figure~\ref{fig:crossproject} shows the distribution of predictive performance on test projects not seen during training. Not unexpectedly, the  performance decreases compared to the intra-project setting -- where training is performed on historical data of the projects -- from 0.58 to 0.36, on average (with Precision@k ranging from 0.43 at $k=10$ to 0.30 at $k=3$).

\vspace{1ex}
{\centering\includegraphics[width=0.45\linewidth]{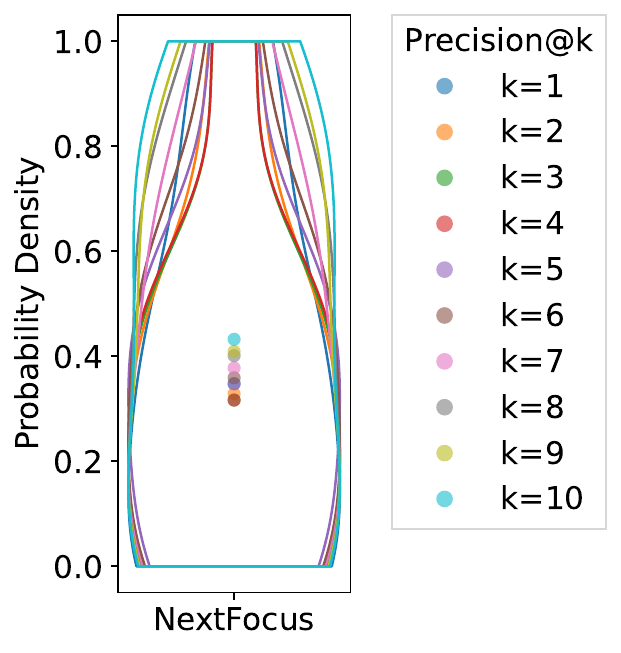}
  \captionsetup{font=small, skip=0pt}
  \captionof{figure}{Distribution of average Precision@$k$ per project in the cross-project setting.}
  \label{fig:crossproject}}
  \vspace{1ex}

\paragraph{Experiment 6.} 
We compare \nameapproach{} against a single-location, state of the art model completion approach, \nametinnes{}.}{}

%
\changedFinalRevision{
We report in Table \ref{tbl:correctnesscomparedtobaseline} correctness across projects, averaged per project, following the same procedure as in Experiment 3 and 5, since results are paired across approaches. Notably, \nameapproach{} achieves significant higher correctness for \textit{next focus prediction} (\CorrectnessValueNextFocus) than \nametinnes (\CorrectnessValueNextFocusTinnes) (Wilcoxon signed-rank test, $p < 0.05$).

}{}

\changedFinalRevision{To isolate the contribution of {\nameapproach}, we 
compare {\nameapproach}+\nametinnes and \nametinnes: no significant difference is observed for 
 \textit{change structure correctness}, and \textit{type structure correctness} ($p > 0.05$) but \nameapproach already achieves significantly higher \textit{structure correctness} and \textit{format correctness} 
 ($p < 0.05$). As outlined in the experimental setup (Section \ref{experiment6}), \nametinnes was not originally designed for iterative multi-location completion, which limits its performance in {\nameapproach}+\nametinnes{}.

As a result, the improved multi-location model completion approach, {\nameapproach}+\nametinnesbetter, 
significantly outperforms the baseline in \textit{next focus prediction}, \textit{change structure}, 
and \textit{structure correctness}.
}{}
\begin{table}[htbp]
\captionsetup{font=small, skip=4pt}
\caption{Average correctness across projects for next focus node prediction and single-location model completion (in~\%).}
\small
\begin{tabular}{p{2.64cm} p{0.7cm} p{0.8cm} p{1.1cm} p{1.1cm} p{1cm}}

\toprule
         &     Next      &          &   
& Change        &  Type  \\

Approach &  Focus & Format & Structure &  Structure &  Structure  \\ 
\midrule
\nametinnes   &  30.18  & 95.63& 16.35& 
13.98& 12.19       \\
{\nameapproach}+\nametinnes & 63.94& 99.67&22.85  & 10.48 & 8.48  \\
{\nameapproach}+\nametinnesbetter & 60.33& 96.12 & 40.68&21.31 & 16.11\\
\bottomrule
\end{tabular}
\label{tbl:correctnesscomparedtobaseline}
\end{table}

%% file: Content/8_Dicussion.tex
\subsection{Discussion}

In a large, real-world software model, a local change can affect other (distant) parts of the model, even if it is well-structured~\cite{rashid2011aspect,apel2010calculus,apel2011language}. To support modelers in finding the relevant locations to change, we propose \nameapproach{} for multi-location model completion. \nameapproach learns co-change patterns from historical data and suggests additional model locations to change.

\paragraph{RQ1.}

Our initial objective was to assess to what extent \nameapproach can predict focus nodes for multi-location software model completion; to this end, we trained, evaluated, and compared \nameapproach against three baselines, \Semantic{}, \Historical{}, \Random{}.
We found that \nameapproach consistently outperformed the baselines, achieving an average score of $0.98$ over all $k \leq 10$, and performed well independently of the number of recommendations.
\nameapproach successfully learns patterns from history, outperforming \Historical{} by better capturing contextual semantics. While \Historical{} alone is insufficient -- since the same type of change can occur in other elements of the same or different models -- semantic embeddings help identify such cases.

At the same time, learning from historical data proves to be effective, as witnessed by \nameapproach superior performance compared to static \Semantic{} alone. By including a random baseline, we verified that the performance is not due to chance. Given the different pattern characteristics -- some involving a large number of changes, others only very few -- \nameapproach consistently predicted relevant nodes, even when only a small number of correct nodes needed to be identified from a large candidate set. Our approach reliably suggests relevant new focus nodes, matching patterns that were actually made by modelers in real-world settings.

\paragraph{RQ2.}

We investigated to what extent \nameapproach{}'s ability to predict new focus nodes depends on the distance between the predicted nodes and the anchor node, examining whether \nameapproach{} can predict both local and global next focus nodes. We found that, while absolute performance slightly decreases with an increasing radius, this trend was to be expected, as more nodes become candidates and the probability of a node being a correct change node decreases. This illustrates how the task becomes harder for \nameapproach in distinguishing relevant from irrelevant nodes. Nevertheless, \nameapproach keeps predictive performance high, even at longer graph distances.
The strong performance, independent of the distance, may be due to the model not relying on graph connections but instead focusing on semantic embeddings and historical co-changes.

\paragraph{RQ3.}

We were interested in the project-specific and pattern-specific properties that influence the performance of \nameapproach. While machine learning performance often depends on factors like training set size and label distribution, we observed no correlation with dataset size -- some large datasets performed poorly, and some small datasets yielded perfect predictions (Figure \ref{fig:figtotalvalues}, green dots). This suggests that even small projects with a few examples may have benefited from other projects. 
The slight performance increase for \Semantic{} may result from datasets with more positives in training also having more in testing, which raises the chance of correct focus nodes with similar semantic embeddings, despite the lack of training.
Overall, performance varied more with the nature of the change pattern:  \nameapproach performs well on recurring patterns such as type replacements, likely because they appear frequently in training and exhibit clear semantic cues. However, \nameapproach predictive performance was lower for uncommon patterns and hierarchy-related changes, though it still generally surpasses the baselines. This may be due to the fact that such cases require an understanding of deeper structural context than \nameapproach provides.
\changedFinalRevision{
Applying \nameapproach on projects that were unseen during training, accounting for cases where historical data may not always be available in real-world scenarios, we observe a drop in performance, which indicates that project-specific historical data indeed helps in making more accurate predictions. Nevertheless, \nameapproach is still able to leverage information from other projects to make reasonable predictions on unseen data (e.g. 0.43 Precision@10).
}{}
\changedFinalRevision{%
\paragraph{RQ4.}

To assess how well \nameapproach supports iterative, multi-location model completion in a realistic workflow, we first compare \nameapproach' capabilities for next focus node prediction against state of the art single-location completion, which is iteratively executed.
Notably, \nameapproach achieves \CorrectnessValueNextFocus correctness compared to \CorrectnessValueNextFocusTinnes for \nametinnes (see Table~\ref{tbl:correctnesscomparedtobaseline}).
We additionally report correctness at multiple levels following the metrics by Tinnes et al.~\cite{tinnes2024leveraging}. As \nametinnes does not support iteratively completing the model $N$ times, a drop in model completion correctness occurs compared to the values stated in the work by~\citet{tinnes2024leveraging}. More specifically, \nametinnes, does not support jumping to new focus nodes in the model due to the edge-only linearization (\textsc{EdgeL} format). The improved version, {\nameapproach}+\nametinnesbetter, however, archives higher correctness,  even in an iterative setting with $N=10$ (e.g.
\StructureImproved \textit{structure correctness}).
}{}

\subsection{Threats to Validity}

With regard to internal validity, our evaluation relies on noisy historical commit data, performed by modelers in real-world scenarios, which, in some cases, may include unrelated or tool-generated changes from EMF; for example, our manual analysis revealed three commits for which it was unclear whether the multi-location changes were conceptually related or simply co-occurred in the same commit by coincidence.
Additionally, the overall differences between modeling projects sizes can influence the overall performance, since bigger projects may dominate the performance. This is exactly why we conducted Experiment 3, which confirmed that \nameapproach consistently outperforms the baselines across projects.

With regard to external validity, we cannot claim transferability to all domains. However, the inclusion of 32 real-world, diverse, open-source modeling projects, with multi-location changes that have been performed by modelers in the real-world, provides strong evidence for the generalizability of our findings. The lack of further publicly available datasets currently hinders the extension of our evaluation on more data~\cite{burgueno2025automation, tinnes2024leveraging, lopez2022modelset, muttillo2024towards}. 
Due to the inherent structure of commit histories (see Section~\ref{experiment1}) and the need for manual semantic analysis, we limited the evaluation to one multi-location pattern per project -- specifically, the most recent one. While this restriction was necessary for manual investigation, future work shall explore earlier versions of the model histories by shifting the train-test split toward older commits.
\changedFinalRevision{For multi-location completion, we partially reimplemented the approach by Tinnes et al.~\cite{tinnes2024leveraging} and acknowledge possible minor deviations from the original.}{}

%% file: Content/7_conclusion.tex
\section{Conclusion and Future Work}

Software models often grow large and complex, undergoing thousands of changes through evolution, refactoring, and maintenance.
With the rise of LLMs, new possibilities have opened up in the software modeling domain. While recent approaches support single-location model completion, we aim to extend this setting to multi-location model completion by proposing \nameapproach{}.
It consists of a node embedding mechanism, an attention-based neural network, and a ranking system. \nameapproach achieves promising results for multi-location model completion, even when changes are largely spread across the model. \nameapproach significantly outperforms the baselines: \Random{}, \Semantic{}, \Historical{}, which reflect concepts common in similar domains~\cite{kagdi2013integrating, zimmermann2005mining, jiang2021investigating, kchaou2017uml, agt2018domore, agt2019automated, lopez2023word, elkamel2016uml, burgueno2021nlp}.
\nameapproach{} excelled in scenarios with recurring change patterns and also performed well on some model extensions. However, its performance was lower for less common patterns and hierarchy-related changes. \changedFinalRevision{\nameapproach benefits from project-specific historical data; 
however, if such data is not available, it can still make use of information 
from other projects to perform reasonably in cross-project settings.
\noindent Finally, combining \nameapproach with single-location completion enables effective iterative, multi-location model completion, achieving \CorrectnessValueNextFocus next focus node correctness.

}{}

\section{Data Availability}
\label{sec:data_availability}
We provide the data and Python code for \nameapproach{} as well the baselines in \appendixsite{}, including training and evaluation scripts to reproduce our analysis.

\newpage